# Potential molecular semiconductor devices: cyclo-Cn (n = 10 and 14) with higher stabilities and aromaticities than acknowledged cyclo-C18

Meng-Yang Li,[a] Zhibin Gao,[b] Yan-Bo Han,[a] Yao-Xiao Zhao,[a] Kun Yuan,[c] Shigeru Nagase,[d] Masahiro Ehara,[e] Xiang Zhao[a*]

[a] *Institute for Chemical Physics & Department of Chemistry, School of Science, State Key Laboratory of Electrical Insulation and Power Equipment & MOE Key Laboratory for Nonequilibrium Synthesis and Modulation of Condensed Matter, Xi'an Jiaotong University, Xi'an710049, China.*

[b] Department of Physics, National University of Singapore, Singapore 117551, Republic of Singapore.

[c] College of Chemical Engineering and Technology, Tianshui Normal University, Tianshui, 741001, China.

[d] Fukui Institute for Fundamental Chemistry Kyoto University, Kyoto 606-8103, Japan.

[e] Institute for Molecular Science, Okazaki, 444-8585, Japan.

Corresponding author *E-mail: xzhao@mail.xjtu.edu.cn

**Abstract**

The successful isolation of cyclo-$C_{18}$ in experiment means the ground-breaking epoch of carbon rings. Herein, we studied the thermodynamic stabilities of cyclo-$C_n$ ($4 \leq n \leq 34$) with density functional theory. When n = 4N + 2 (N is integer), cyclo-$C_n$ ($10 \leq n \leq 34$) were thermodynamically stable. Especially, cyclo-$C_{10}$ and cyclo-$C_{14}$ were thermodynamically, kinetically, dynamically, and optically preferred to cyclo-$C_{18}$, and were the candidates of zero-dimension carbon rings. The carbon atoms were *sp* hybridization in cyclo-$C_{10}$, cyclo-$C_{14}$ and cyclo-$C_{18}$. Cyclo-$C_{14}$ and cyclo-$C_{18}$ had alternating abnormal single and triple bonds, but cyclo-$C_{10}$ had equal bonds. Cyclo-$C_{10}$, cyclo-$C_{14}$ and cyclo-$C_{18}$ with large aromaticities had out-plane and in-plane π systems, which were perpendicular to each other. The number of π electrons in out-plane and in-plane π systems followed Hückel rule. Simulated UV-vis-NIR spectra indicated the similar electronic structures of cyclo-$C_{14}$ and cyclo-$C_{18}$.

**Introduction**

The carbon atom attracts a great deal of attentions because it is a fundamental atom in organic compounds and is the constituent atom for many useful materials. For example, there are some pure carbon materials, including zero-dimension fullerenes,[1] one-dimension nanotube,[2] two-dimension graphene,[3] and three-dimension diamond[4] and graphite. They are well-known as the allotropes of carbon atom and are useful in many fields due to their unique electronic structures and physical chemical features. In the past, fullerene is the only zero-dimension allotrope of carbon.

Very recently, the successful synthesis and isolation of cyclo-$C_{18}$ are epoch-making in the carbon chemistry and mean a second zero-dimension allotrope of carbon atom.[5]

In fact, the early experiments on pure carbon clusters date back to 1954 using a heated substrate as cluster source.[6] After that, there are several experimental and theoretical studies on the cyclo-$C_n$. The synthesis and x-ray crystal structure of a direct precursor of cyclo-$C_{18}$ were described in 1989. The results of time-of-flight mass spectroscopy recognized cyclo-$C_{18}$ as the predominant fragmentation pattern with the flash heating experiments on this precursor.[7] In addition, Tobe and Diederich *et al.* synthesized a number of precursors of cyclo-$C_{18}$ with well-defined cyclic geometries and readily removable groups. These precursors eventually generated cyclo-$C_{18}$ by laser desorption induced [4 + 2] cycloreversion (retro-Diels-Alder reaction), decarbonylation or [2 + 2] cycloreversion.[8, 9]. Rubin revealed that cyclo-$C_{18}$ can be highly stabilized as a ligand in complex characterized by X-ray crystallography.[10] Furthermore, the mass spectral evidences of $C_{18}$ are clearly detected during studying the fullerene-formation mechanism.[11] For example, the coalescence of cyclo-$C_{30}$ predominantly produces buckminsterfullerene ($C_{60}$), and the small rings cyclo-$C_{18}$ and cyclo-$C_{24}$ preferentially produce fullerene $C_{70}$ through distinct intermediates.[11] The electronic spectra of $C_{18}$ and $C_{22}$ were detected in the gas phase by mass-selective resonant two-color two-photo ionization technique coupled to a laser ablation source, and the absorption spectrum of $C_{14}$ was also observed.[12] Another method to generate the small carbon cluster ions is presented via reaction with HCN.[13] Some of the detected cyanopolyynes were also observed in the interstellar medium. Circumstellar carbon condensation processes in the atmospheres of carbon-rich stars, which are similar to those studied in the laboratory, are suggested as possible synthetic sources.[14-17] Although there are many evidences for the cyclo-$C_n$ clusters, the geometries of cyclo-$C_n$ are not determined in experiment up to now.

The previous theoretical investigation contribute much to studying the geometries of cyclo-$C_n$.[18-21] However, there are lots of controversies on the alternating bonds or equal bonds for cyclo-$C_n$ from the viewpoints of experiment and theory. In the case that $N \leq 4$, the bond-length nonalternating cumulenic structure [$D_{(2N+1)h}$] for monocyclic carbon $4N + 2$ is found to be the most stable, but this structure becomes unstable when $N \geq 5$.[22] Then the bond-length alternant structure [$C_{(2N+1)h}$] becomes the most stable among the ring-shape clusters. However, the results of the local-density approximation show that rings $C_{4N+2}$ present no alternation (~i.e., aromatic behavior is retained) until very large sizes ($N > 20$).[23] Clearly, the theoretical results cannot be confirmed because of no exactly experimental data on rings $C_{4N+2}$.

After the successful isolation and characterization of cyclo-$C_{18}$ with high-resolution atomic microscopy, the alternating bonds for cyclo-$C_{18}$ are clearly revealed.[5] During these previous studies on carbon cluster, the other cyclo-$C_n$ were also detected in experiments.[24, 25] The isolation of cyclo-$C_{18}$ encourages both experimental and theoretical researchers to explore the stabilities and physic-chemical properties of cyclo-$C_n$. In the similar case of polyacetylene, previous studies reported that the HF (LDA and GGA) method tends to overestimate (underestimate) the degree of bond-length alternation in polyacetylene, while the hybrid density functional theory (DFT) as well as the Möller-Plesset second-order perturbation theory well reproduces the experimental degree.[22] Luckily, there is also the benchmark of cyclo-$C_{18}$ to select the suitable theoretical method. Herein, we explored the stabilities of cyclo-$C_n$ (4 ≤ n ≤ 34) with hybrid DFT,[26] and the cyclo-$C_n$ (10 ≤ n = 4$N$+2 ≤ 34, N is integer) would be the thermodynamically stable molecules. Especially, cyclo-$C_n$ (n = 10 and 14) with two similarly perpendicular π systems, including in-plane and conventional out-plane π systems, had larger aromaticities and higher stabilities than the acknowledged cyclo-$C_{18}$.

**Calculation Methods**

The optimization of cyclo-$C_{18}$ was carried out with the B3LYP/6-31G(d, p) and M062X/6-311G(d, p) (Figure S1).[27-29] The results on the M062X/6-311G(d, p) are well consistent with the experimental results, possessing the similar alternating bond lengths (1.35/1.34 Å and 1.23/1.20 Å).[5] Thus, we optimized cyclo-$C_n$ (4 ≤ n ≤ 34) with the M062X/6-311G(d, p), and the optimizations of cyclo-$C_n$ (4 ≤ n ≤ 34) were without any imaginary frequency on the M062X/6-311G(d, p). The thermodynamically optimal cyclo-$C_n$ (10 ≤ n = 4$N$+2 ≤ 34, $N$ is integer), with lower energies per carbon atom than adjacent atoms, attracted our great interest, especially cyclo-$C_{10}$, cyclo-$C_{14}$, and cyclo-$C_{18}$. In order to clarify the electronic structures of cyclo-$C_n$ (n = 10, 14, and 18), we carried out the natural bond orbital analysis and UV-vis-NIR spectra on the M062X/6-311G(d, p) level.[30,31] The aromacities of cyclo-$C_n$ (n = 10, 14, and 18) were evaluated with nucleus-independent chemical shifts (NICS),[32-34] a simple and efficient aromaticity probe. All of above calculations were carried out with Gaussian 16 program package.[35]

In addition, the standard *ab* initio molecular dynamic, in which a Verlet algorithm was used to integrate Newton's equations of motion, was simulated on the Perdew-Burke-Ernzerhof (PBE) functional-style generalized gradient approximation (GGA) for the optimal cyclo-$C_n$ (n = 10, 14, and 18) under microcanonical (NVE) ensemble with the VASP program package.[36–39] The thermodynamically stable cyclo-$C_n$ (n = 10, 14, and 18) were optimized again with the VASP to obtain the initial positions of molecular dynamic calculation. The default cut-off energy for the pseudopotentials was 400 eV, and the tetragonal lattices were 20 Å for all systems. The three dimensional periodic boundary condition with a minimum inter-molecular distance of 12 Å was used for all cases to ensure the negligible intermolecular interaction. Besides, an automatic k-points mesh (1 × 1 × 1) was generated using the Gamma centered grid. The time step was one in femtosecond, and the constant temperature considered was 300 K.

**Results and Discussion**

**The prior stabilities of cyclo-$C_n$ (n = 10 and 14) to acknowledged cyclo-$C_{18}$**

The optimization results of cyclo-$C_n$ (4 ≤ n ≤ 34) with different spin multiplicity on the M062X/6-311G(d, p) level are shown in Table S1. As shown in Table S1, cyclo-$C_n$ (13 ≤ n ≤ 34) with odd carbon atoms have triplet ground states, and the other cyclo-$C_n$ (4 ≤ n ≤ 34) have singlet ground states. The singlet cyclo-$C_n$ (n = 5, 7, and 9) were non-plane. The cyclo-$C_{11}$ had singlet ground state instead of triplet because the relative small ring had well delocalized electrons and close annulene. The spin density maps of triplet cyclo-$C_n$ (n = 2m + 1, m = 6-16) are shown in Figure S2. It is clear that triplet cyclo-$C_n$ (n = 2m + 1, m = 6-16) consist of annulene and alkynyl. The spin electrons almost localize in annulene part. The cyclo-$C_n$ (4 ≤ n ≤ 34) with triplet ground state are not stable because of their radical characters.[40, 41] Therefore we did not consider much on these cyclo-$C_n$ (4 ≤ n ≤ 34) with triplet ground states.

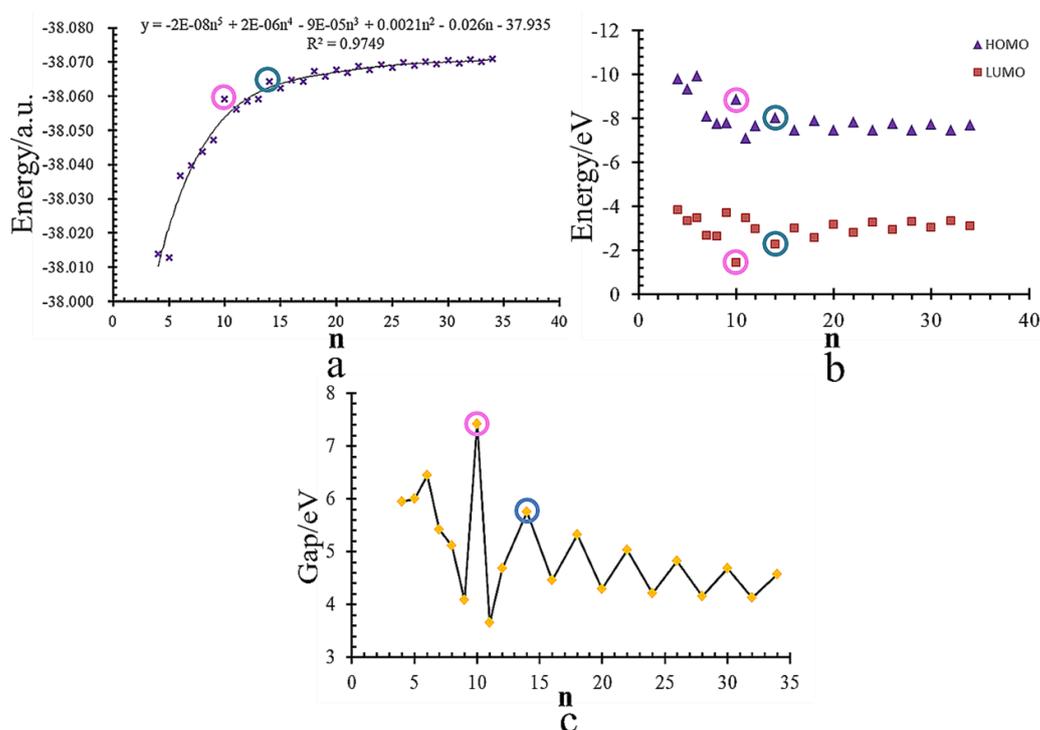

Figure 1 a) The energies per carbon atom of cyclo-$C_n$ (4 ≤ n ≤ 34), b) the HOMO and LUMO energy level and c) HOMO-LUMO gap for cyclo-$C_n$ (4 ≤ n ≤ 34) with singlet ground state based on the M062X/6-311G(d,p). Cyclo-$C_{10}$ and cyclo-$C_{14}$ marked in pink and blue circle, respectively.

The thermodynamic stabilities of cyclo-$C_n$ (4 ≤ n ≤ 34) were explored with the energy per carbon atom in the optimal cyclo-$C_n$ (4 ≤ n ≤ 34) with ground states. This method has been confirmed in exploring the thermodynamic stability of another allotrope of carbon, fullerenes, and is also carried out to clarify the reason why $C_{60}$ and $C_{70}$ possess the largest yield among fullerenes.[42,43] As shown in Figure 1a, each carbon atom in cyclo-$C_n$ ($10 \leq n = 4N + 2 \leq 34$, $N$ is integer) possesses lower energies than it in their adjacent rings, meaning that cyclo-$C_n$ ($10 \leq n = 4N + 2 \leq 34$) were thermodynamically stable molecules. With the increasing n, the energies per carbon atoms of cyclo-$C_n$ ($4 \leq n \leq 34$) converged to a constant. Cyclo-$C_{10}$ and cyclo-$C_{14}$ had lower energies per carbon atoms than their respective adjacent homologue. The gaps between energies per carbon atom of cyclo-$C_{10}$ and cyclo-$C_{14}$ and their homologues were larger than that of cyclo-$C_{18}$. Thus, the cyclo-$C_{10}$ and cyclo-$C_{14}$ were more thermodynamically stable than cyclo-$C_{18}$. These findings suggested that the thermodynamically stable cyclo-$C_n$ ($10 \leq n \leq 34$) followed the n = $4N + 2$ rule.

Within the validity of Koopmans' theorem, the energy of highest occupied molecular orbital (HOMO) is the negative ionization potential.[44,45] The lower energy of HOMO is, the more difficult ionization is for molecules. As shown in Figure 1b, when n = $4N + 2$, the cyclo-$C_n$ ($4 \leq n \leq 34$) with singlet ground states possess lower HOMO energies than their adjacent molecules. Especially, the cyclo-$C_{10}$ and cyclo-$C_{14}$ have lower HOMO level than cyclo-$C_{18}$. In addition, the energies of the lowest unoccupied molecular orbitals (LUMOs) of the cyclo-$C_{10}$ and cyclo-$C_{14}$ are higher than that of cyclo-$C_{18}$ LUMO. Higher LUMO energy means that it is more difficult to obtain electrons for neutral molecules. Based on the above discussions on HOMOs and LUMOs, the cyclo-$C_{10}$ and cyclo-$C_{14}$ are more chemically inert than cyclo-$C_{18}$.

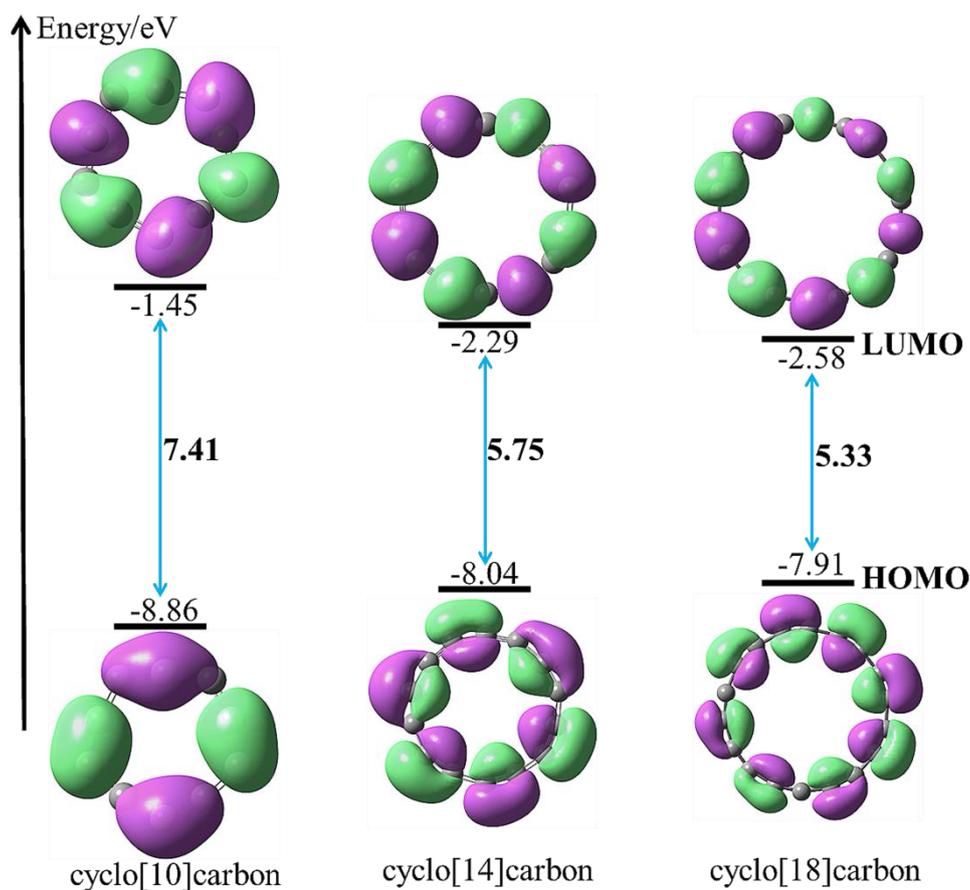

Figure 2 The distribution of HOMO and LUMO for cyclo-$C_n$ (n = 10, 14 and 18) based on the M062X/6-311G(d,p)..

Meanwhile, the HOMO-LUMO gaps for singlet cyclo-$C_n$ (4 ≤ n ≤ 34) also followed the $4N + 2$ rule. When n = $4N + 2$ for cyclo-$C_n$ (4 ≤ n ≤ 34), it is clear that the HOMO-LUMO gaps (Figure 1c) decrease with the increasing number of carbon atoms and eventually converge to a constant. Especially, cyclo-$C_{10}$ and cyclo-$C_{14}$ possess much larger HOMO-LUMO gaps, and Figure 2 shows their HOMOs and LUMOs. In the viewpoint of optical polarizability, a small HOMO-LUMO gap means small excitation energies to manifold of excited states.[46] The much larger HOMO-LUMO gap for cyclo-$C_{10}$ can be explained with the distributions of its HOMO, which is different from the HOMOs of cyclo-$C_{14}$ and cyclo-$C_{18}$. In fact, the geometry of cyclo-$C_{10}$ was non-alternation ring distinguished from alternating bonds in cyclo-$C_{14}$ and cyclo-$C_{18}$. This similar result, the cumulenic molecule with larger HOMO-LUMO *gap* than acetylenic molecules, also occurs in the linear $C_n$ clusters.[47,48] In a conclusion, cyclo-$C_{10}$ and cyclo-$C_{14}$ were more inert to the illumination than cyclo-$C_{18}$.

In order to further compare the stabilities of cyclo-$C_n$ ( n = 10, 14, and 18), we considered the dynamic stability that the stabilities of molecules are evaluated via the amplitude of energies, bond lengths or bond angles with the changing time.[49] Here, we considered the amplitude of energies at different time scale as the function of stability. As shown in Figure 3, the dynamic study shows that the energy differences between the maximum energy and the minimum one are 0.99, 1.18, and 1.47 a.u. for cyclo-$C_n$ (n = 10, 14, and 18) after 0.5 ps, respectively. Thus, the dynamic stabilities of cyclo-$C_n$ (n = 10 and 14) are superior to the acknowledged cyclo-$C_{18}$. In another word, the geometries of cyclo-$C_n$ (n = 10 and 14) were more difficult to deviate from their optimal geometries than cyclo-$C_{18}$.

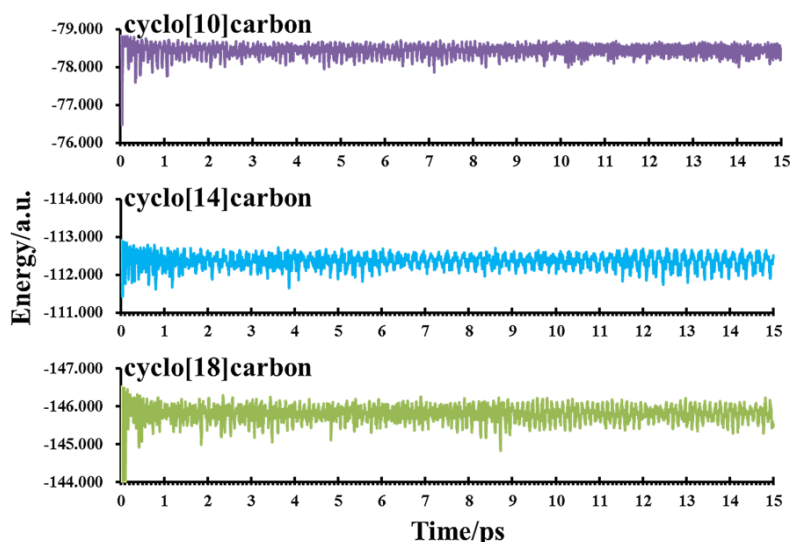

Figure 3 The energies of cyclo-$C_n$ (n = 10, 14 and 18) at different time.

In conclusion, the thermodynamic stabilities for cyclo-$C_n$ ($4 \leq n \leq 34$) follow the $n = 4N+2$ ($N$ is integer) rule when n > 6. Especially, the thermodynamic stabilities, chemically kinetic stabilities, photostabilities, and dynamic stabilities of cyclo-$C_n$ (n = 10 and 14) were superior to those of the acknowledged cyclo-$C_{18}$. In addition, the sequence of stabilities for cyclo-$C_n$ (n = 10, 14, and 18) followed cyclo-$C_{10}$ > cyclo-$C_{14}$ > cyclo-$C_{18}$. Thereby, it is conservative to theoretically predict that it is possible to synthesize and isolate cyclo-$C_{10}$ and cyclo-$C_{14}$ following the isolation of the first carbon ring cyclo-$C_{18}$. Furthermore, besides the recent method to prepare and isolate cyclo-$C_{18}$, there are also several considerable methods, including flash heating and laser vaporization, to prepare the precursors of cyclo-$C_{18}$, oxide and metal complex,.[7-9] All of these methods can be good choices for trying to prepare cyclo-$C_{10}$ and cyclo-$C_{14}$. This prediction is consistent with the previous point that cyclo-$C_{14}$ would be the first carbon ring, with extended Hückel calculations.[18]

**The geometries of cyclo-$C_n$ (n = 10, 14 and 18)**

The geometries of cyclo-$C_n$ (n = 10, 14 and 18) are shown in Figure 4, and their geometrical parameters, including bond lengths and bond angles, are listed in Table

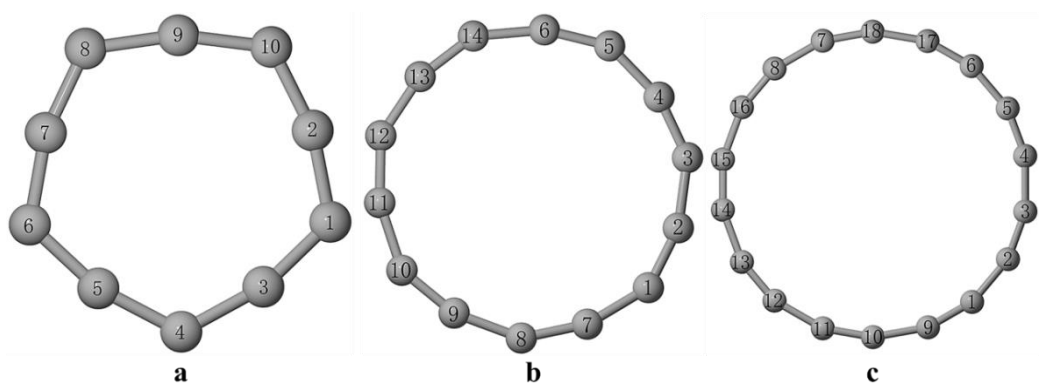

Figure 4 The geometries of a) cyclo-$C_{10}$, b) cyclo-$C_{14}$, and c) cyclo-$C_{18}$ on the M062X/6-311G(d,p).

S2. The maximum distances between two arbitrary atoms are almost constants, 4.14/4.11 Å for cyclo-$C_{10}$, 5.76 Å for cyclo-$C_{14}$, and 7.40 Å for cyclo-$C_{18}$. All of these maximum distances in cyclo-$C_n$ (n = 10, 14, and 18) are close to the diameter of rings. In general, therefore, it means that the geometries of cyclo-$C_n$ (n = 10, 14, and 18) are close to the standard circle. The bond lengths in cyclo-$C_{10}$ are equal, and annulene of cyclo-$C_{10}$ contributed to the larger HOMO-LUMO gap. The alternating bonds, including abnormal single bonds and triple bonds, occur in cyclo-$C_{14}$ and cyclo-$C_{18}$. The triple bonds in cyclo-$C_{14}$ and cyclo-$C_{18}$ are 1.25 Å and 1.23 Å, respectively, which are a little longer than normal C≡C triple bond (1.20 Å). However, the single bonds (1.32 Å and 1.35 Å) in cyclo-$C_{14}$ and cyclo-$C_{18}$ are shorter than normal C-C single bond (1.50 Å). The abnormal single and triple bonds in cyclo-$C_{14}$ and cyclo-$C_{18}$ mean the much contribution of *s* shell of atomic orbitals. Furthermore, the abnormal double bonds in cyclo-$C_{10}$ and abnormal single and triple bonds in cyclo-$C_{14}$ and cyclo-$C_{18}$ could be confirmed with bond orders.[50] The Mayer bond orders (MBOs) between two adjacent carbon atoms in cyclo-$C_{10}$ are 1.78, smaller than 2. Similarly, the MBOs in cyclo-$C_{14}$/cyclo-$C_{18}$ are 1.48/1.28 and 2.14/2.45 for abnormal single and triple bonds, respectively. In cyclo-$C_{14}$ and cyclo-$C_{18}$, the MBOs for abnormal single and triple bonds are clearly larger than 1 and are smaller than 3, respectively. The alternate bond orders also indicated the alternating bonds in cyclo-$C_{14}$ and cyclo-$C_{18}$.

The distortions of molecules can be defined by the difference between ideal angles based on different hybridization of carbon atoms and real angles in target molecules. Carbon atoms in cyclo-$C_{10}$, cyclo-$C_{14}$ and cyclo-$C_{18}$ performed as *sp* hybridization, which was revealed via natural bond orbital analysis. The ideal angle is 180° for *sp* carbon atoms. The angles in cyclo-$C_n$ (n = 10, 14, and 18) are alternative. Thus, we defined the average angles as real angles. The average angles are 144.00°, 154.28° and 160.00° for cyclo-$C_n$ (n = 10, 14, and 18), respectively. Thus, the distortions of cyclo-$C_n$ (n = 10, 14, and 18) are 36.0°, 25.7°, and 20.0°, respectively. The distortions of cyclo-$C_n$ (14 and 18) are similar. Notably, the distortion of cyclo-$C_{10}$, performing as annulene with equal bond lengths, is larger than that of cyclo-$C_n$ (14 and 18).

**The electronic structures and aromaticity of cyclo-$C_n$ (n = 10, 14, and 18)**

The natural bond orbital (NBO) results showed that the carbon atoms were *sp* hybridization in cyclo-$C_n$ (n = 10, 14, and 18), thus each carbon atom had two perpendicular *p* orbitals. In addition, the nucleus-independent chemical shifts (NICS) of cyclo-$C_n$ (n = 10, 14, and 18) are shown in Table 1. Cyclo-$C_{14}$ has the smallest NICS (-32.12) and NICS(1)_ZZ (-74.89) followed by cyclo-$C_{10}$ with NICS (-29.45) and NICS(1)_ZZ (-57.71). Clearly, the aromaticity of cyclo-$C_{14}$ is the largest, and the

Table 1 The nucleus-independent chemical shifts for cyclo-$C_n$ (10, 14, and 18).

|  | NICS[a] | NICS(1)_ZZ[b] |
|---|---|---|
| cyclo-$C_{18}$ | -16.67 | -40.98 |
| cyclo-$C_{14}$ | -32.12 | -74.89 |
| cyclo-$C_{10}$ | -29.45 | -57.71 |

[a] Based on the center of mass for rings;[32] [b] based on the π contribution to the out-of-plane *zz* tensor component.[33]

aromaticity of cyclo-$C_{18}$ is the smallest. It is well known that the aromaticity of molecule is related to molecular stability.[51, 52] Molecules with larger aromaticity hold higher stability. The results of aromaticity again confirmed that the stabilities of

cyclo-$C_{14}$ and cyclo-$C_{10}$ were higher than that of cyclo-$C_{18}$.

Based on the analysis of NBO and aromaticity of cyclo-$C_n$ (n = 10, 14, and 18), it can be deduced that there were large delocalized π electrons on cyclo-$C_n$ (n = 10, 14, and 18). Here, we studied the π systems of cyclo-$C_n$ (n = 10, 14, and 18) with molecular orbitals and localized orbital locator (LOL) via Multiwfn.[53,54] All of π orbitals of cyclo-$C_n$ (n = 10, 14, and 18) are shown in Figure S3-S5. As shown in

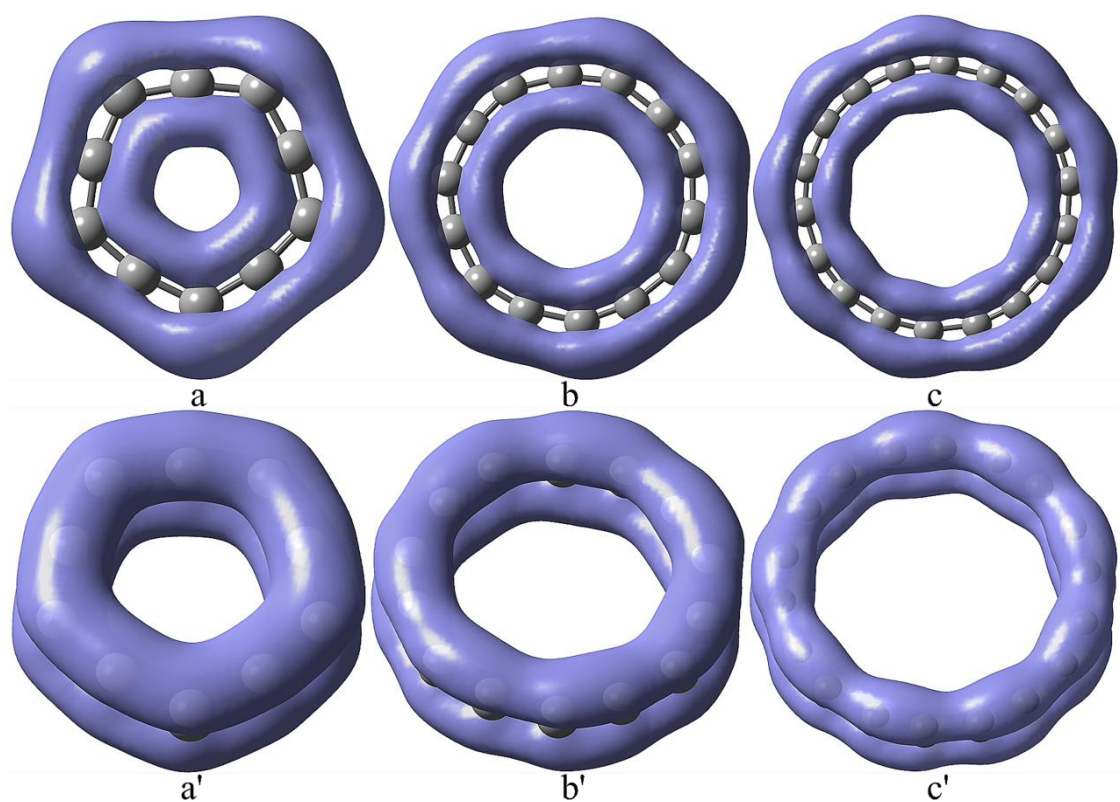

Figure 5 The localized molecular orbitals for cyclo-$C_{10}$ (a and a'), cyclo-$C_{14}$ (b and b') and cyclo-$C_{18}$ (c and c').

Figure S3-S5, all of the number of π orbitals is equal to the number of carbon atoms in cyclo-$C_n$ (n = 10, 14, and 18). These π orbitals can be divided into two types, including in-plane and out-plane π orbitals because of two perpendicular p orbitals on each carbon atoms in cyclo-$C_n$ (n = 10, 14, and 18).

In order to further insight into the π electrons, the localized molecular π orbitals of cyclo-$C_n$ (n = 10, 14, and 18) are shown in Figure 5. The π orbitals, including in-plane and out-plane π orbitals, of cyclo-$C_n$ (n = 10, 14, and 18) were similar to each other. The results of π-LOL in Figure 5 reveal that the π electrons of cyclo-$C_n$ (n = 10, 14, and 18) delocalize on the whole molecules. Similarly, two perpendicular π systems with 4n + 2 π electrons, respectively, follow the Hückel rules. The number of π electrons can be calculated with the number of π molecular orbitals (Figure S3-S5) or the number of electrons on *sp* carbon atoms. For example, cyclo-$C_{10}$ had five in-plane π orbitals and five out-plane π orbitals. Thus it has 10 electrons on each delocalized π orbitals, and the total number of π electrons is 20. On the other hand, there are 10 *sp* carbon atoms in cyclo-$C_{10}$, and the total number of electrons is 60. Each atom has one nonbonding orbital with 20 electrons and one *sp* σ bonds with 20 electrons. There are also 20 π electrons, which delocalized on two delocalized π orbitals, and each delocalized π orbital in Figure 5a and 5a' has 10 π electrons in cyclo-$C_{10}$. The number of π electrons on out-plane and in-plane delocalized π orbitals, respectively following Hückel 4*N* + 2 rule, is the reason why cyclo-$C_n$ (n = 10, 14, and 18) have large aromaticity.

**The UV-vis-NIR spectra for cyclo-$C_n$ (n = 10, 14, and 18)**

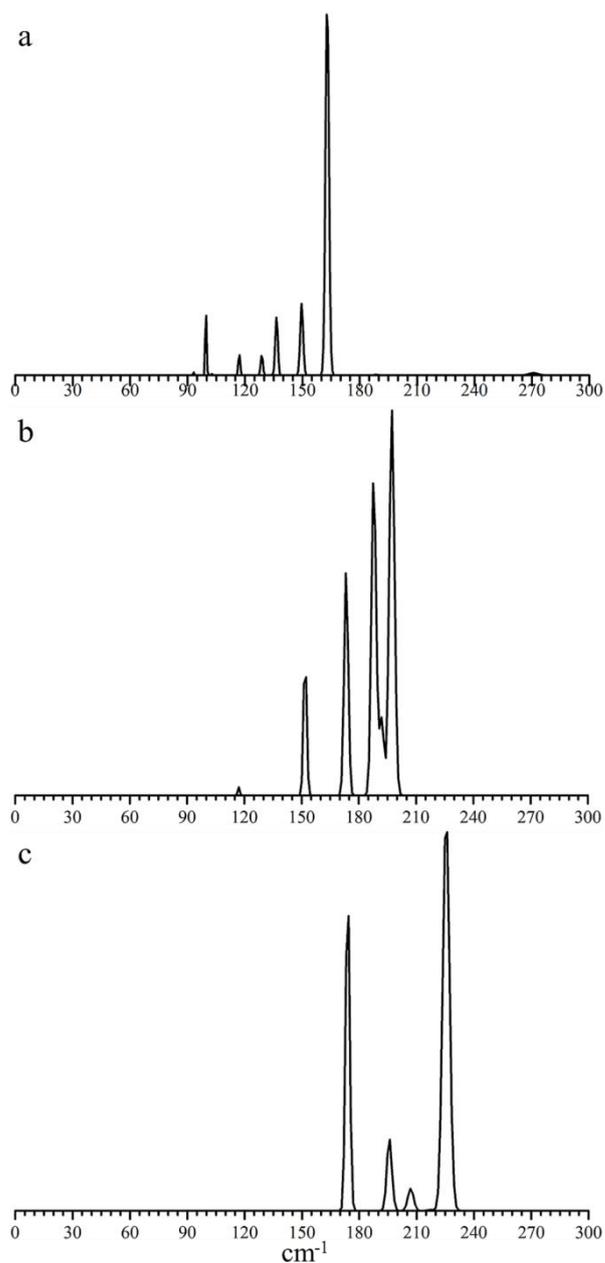

Figure 6 The simulate UV-vis-NIR absorption spectra for cyclo-$C_n$ (n = 10, 14, and 18) on the M062X/6-311G(d,p) level.

The UV-vis-NIR spectrum is an effective method to clarify the electronic structures of organic molecules in experiment.[55-57] Accordingly, we simulated the UV-vis-NIR spectra of cyclo-$C_n$ (n = 10, 14, and 18) on the M062X/6-311G(d,p) level. As shown in Figure 6, the four main absorbed peaks can be observed for cyclo-$C_n$ (n = 10, 14, and 18). Additionally, there are two small absorbed peaks in cyclo-$C_{10}$ distinguished from the cyclo-$C_n$ (n = 14, and 18) with alternating bonds. The absorbed peaks of cyclo-$C_n$ (n = 10, 14, and 18) in Figure 6 are from the π-π* transition. The similar absorbed peaks for cyclo-$C_n$ (n = 14, and 18) represent the similar geometries and electronic structures. With the ring increasing, the blue shifts of four main absorbed peaks in cyclo-$C_n$ (n = 10, 14, and 18) occur due to the more π electrons. The simulated UV-vis-NIR spectra of cyclo-$C_n$ (n = 10, 14, and 18) will be the good assistance to further explore cyclo-$C_n$ (n = 10, 14, and 18) in experiment.

**Conclusion**

Herein, the thermodynamic stabilities of cyclo-$C_n$ (4 ≤ n ≤ 34) were explored with density functional theory, and the results revealed that cyclo-$C_n$ (10 ≤ n ≤ 34, n = $4N + 2$) would be thermodynamically stable, which followed the Hückel $4N + 2$ rule. Compared with the acknowledged cyclo-$C_{18}$, cyclo-$C_{10}$ and cyclo-$C_{14}$ possessed higher thermodynamic, kinetic, optical, and dynamic stabilities. Accordingly, we predicted that cyclo-$C_{10}$ and cyclo-$C_{14}$ would be prepared in future experiments. There were equal double bonds in cyclo-$C_{10}$, as was different from the alternating abnormal single bonds and triple bonds in cyclo-$C_{14}$ and cyclo-$C_{18}$. These geometrical characters were also in accordance with the Mayer bond orders between two adjacent carbon atoms. The cyclo-$C_{18}$, cyclo-$C_{10}$ and cyclo-$C_{14}$ had large aromaticities because there were two perpendicular delocalized π orbitals, including out-plane and in-plane π orbitals, on them. The number of π electrons on the out-plane and in-plane π orbitals followed Hückel $4N + 2$ rule. Furthermore, the much larger aromaticities of cyclo-$C_{10}$ and cyclo-$C_{14}$ are also in favor of the higher stabilities of cyclo-$C_{10}$ and cyclo-$C_{14}$ than

cyclo-$C_{18}$. The UV-vis-NIR spectra of cyclo-$C_n$ (n = 10, 14, and 18) were simulated to be a reference for the further experimental study. There are lots of efforts on preparing cyclo-$C_{18}$, and all of these methods can be carried out for trying to prepare cyclo-$C_n$ (n = 10 and 14) with larger stability than cyclo-$C_{18}$. Undoubtedly, the successful isolation and characterization of cyclo-$C_{18}$ will encourage many researchers to study cyclo-$C_n$ homologues, and based on the present theoretical investigation, we believe that cyclo-$C_n$ (n = 10 and 14) will be the potential candidates for zero-dimension allotrope of carbon atom.

**Acknowledgements**

The National Natural Science Foundation of China (21773181, 21573172) has financially supported this work. X. Z. would like to acknowledge the financial support from the Nanotechnology Platform Program (Molecule and Material Synthesis) of the Ministry of Education, Culture, Sports, Science, and Technology of Japan. Z. Gao acknowledges the financial support from MOE tier 1 funding of NUS Faculty of Science, Singapore (Grant No. R-144-000-402-114).

**Author contributions**

M.Y. Li, Y. X. Zhao and X. Zhao conceived the ideas and designed the project. M.Y.

Li performed DFT calculations and wrote the manuscript. Z. B. Gao, Y. X. Zhao, Y. B. Han, K. Yuan and X. Zhao revised the manuscript. M.Y. Li and Y. X. Zhao contributed equally to this work.

**Competing interests**

The authors declare no competing interests.